\documentclass[12pt, draftclsnofoot, onecolumn]{IEEEtran}
\usepackage{romannum}
\usepackage{graphicx}
\usepackage{multirow}
\usepackage{amssymb,bm}
\usepackage{algpseudocode}
\usepackage{algorithm}
\usepackage{array}
\usepackage{subfigure}
\usepackage{color}
\usepackage[utf8]{inputenc}
\usepackage[T1]{fontenc}
\usepackage{url}
\usepackage{ifthen}
\usepackage{cite}
\usepackage{mathtools}
\usepackage[english]{babel}
\usepackage{amsthm}
\usepackage{thmtools}
\usepackage{enumerate}

\declaretheoremstyle[%
  spaceabove=6pt,%
  spacebelow=6pt,%
  headfont=\normalfont\itshape,%
  bodyfont=\normalfont,%
  postheadspace=1em,%
  qed=\qedsymbol%
]{mystyle}

\declaretheoremstyle[%
  spaceabove=6pt,%
  spacebelow=6pt,%
  headfont=\bfseries,%
  bodyfont=\normalfont,%
  postheadspace=1em,%
]{mystyle_1}

\declaretheorem[name={Proof},style=mystyle,unnumbered,
]{prf}
\declaretheorem[name={Theorem},style=mystyle_1,
]{theo}

\declaretheorem[name={Corollary},style=mystyle_1,
]{corollary}
\declaretheorem[name={Lemma},style=mystyle_1,
]{lemma}
\declaretheorem[name={Definition},style=mystyle_1,
]{defi}

\emergencystretch=\maxdimen
\ifCLASSINFOpdf
\else

\fi

\begin{document}
\pagenumbering{arabic}
\title{A Novel Sum-Product Detection Algorithm for Faster-than-Nyquist Signaling: A Deep Learning Approach}

\author{%
  \IEEEauthorblockN{Bryan~Liu,~\IEEEmembership{Student Member, ~IEEE,}
  Shuangyang Li,~\IEEEmembership{Student Member, ~IEEE,}
  Yixuan Xie,~\IEEEmembership{Member, ~IEEE,}
  and
  Jinhong~Yuan,~\IEEEmembership{Fellow, ~IEEE} \\}
  \thanks{
   B. Liu, S. Li, Y. Xie and J. Yuan are with the School of Electrical Engineering and Telecommunications, the University of New South Wales, Sydney, Australia (e-mail: \{bryan.liu, shuangyang.li, yixuan.xie, j.yuan\}@unsw.edu.au). This paper was presented in part at the 2019 IEEE Information Theory Workshop (ITW) \cite{Bryan_ITW2019}.
  }
}

\maketitle

\vspace{-4mm}
\begin{abstract}
A deep learning assisted sum-product detection algorithm (DL-SPDA) for faster-than-Nyquist (FTN) signaling is proposed in this paper.
The proposed detection algorithm works on a modified factor graph which concatenates a neural network function node to the variable nodes of the conventional FTN factor graph to approach the maximum \emph{a posterior} probabilities (MAP) error performance.
In specific, the neural network performs as a function node in the modified factor graph to deal with the residual intersymbol interference (ISI) that is not considered by the conventional detector with a limited complexity.
We modify the updating rule in the conventional sum-product algorithm so that the neural network assisted detector can be complemented to a Turbo equalization receiver.
Furthermore, we propose a compatible training technique to improve the detection performance of the proposed DL-SPDA with Turbo equalization.
In particular, the neural network is optimized in terms of the mutual information between the transmitted sequence and the extrinsic information.
We also investigate the maximum-likelihood bit error rate (BER) performance of a finite length coded FTN system.
Simulation results show that the error performance of the proposed algorithm approaches the MAP performance, which is consistent with the analytical BER.
\end{abstract}
\IEEEpeerreviewmaketitle

\section{Introduction}
Faster-than-Nyquist (FTN) signaling \cite{FTN_Mazo,Anderson_FTN,Detection-li2018superposition} has long been one of the promising communication paradigms for future high data rate wireless networks.
Different from the conventional methods of enhancing the data rate, which normally requires more time/bandwidth/spatial resources, FTN signaling
enhances the spectral efficiency by intentionally transmitting the symbols faster than the Nyquist rate.
More importantly, FTN signaling is able to achieve  the ultimate capacity for the signal power spectral density (PSD) \cite{Capacity_FTN_1}.
Therefore, FTN signaling has been widely proposed for various communication applications, such as satellite communications \cite{TFP}, and beyond 5G communications \cite{towards2019lee, yuan2020iterative}.

A major drawback of FTN signaling is that the higher symbol rate induces inevitable and severe intersymbol interference (ISI) at the transmitter side. Consequently, a high complexity detector is usually required at the receiver side \cite{Anderson_FTN,TE_First,TE_Journal,yuan2019iterative}. For example, the number of trellis states in a Bahl-Cocke-Jelinek-Raviv (BCJR) detector increases exponentially with the constellation size and number of ISI taps.
Moreover, for a coded FTN system, Turbo equalization is usually applied at the receiver, where iterations are performed between the BCJR detector and channel decoder. The overall detection/decoding complexity further increases with respect to the number of iterations.
Therefore, designing practical detectors with reduced-complexity is a major research topic for FTN signaling \cite{Shuangyang,ReducedMBCJR,MBCJR_Covalope}.
For example, two M-algorithm BCJR (M-BCJR) detectors were proposed for FTN signaling in \cite{Shuangyang} based on the Ungerboeck observation model \cite{Ungerboeck}, and they show promising error performance for coded FTN systems by applying Turbo equalization.
Other than the BCJR algorithm, sum-product algorithms (SPA) have also been recognized as an efficient method to compute the marginal probabilities with a low complexity \cite{Kschischang2019factor} and it has been
widely used in the decoding of channel codes \cite{DE_LDPC} and signal processing \cite{AMP}. A soft-input soft-output (SISO) detection algorithm for linear ISI channels was proposed in \cite{SISO}, where the sum-product algorithm is applied to a factor graph (FG) of the coded FTN system.
The complexity of the algorithm is linear to the number of interferers during each iteration.

However, reduced-complexity detection algorithms usually undermine the error performance.
For instance, there are mainly two different aspects that may contribute to the performance loss if we apply the SISO algorithm in \cite{SISO} for FTN detection.
Firstly, due to the complexity limitation, the SISO algorithm can only consider a limited number of ISI taps.
Note that the number of ISI taps of FTN signaling can be infinite in theory \cite{Anderson_FTN}. Therefore,
the unconsidered residual ISI will degrade the error performance of the system.
Secondly, the FG generated from the Ungerboeck observation model in \cite{SISO} has shortest cycles of length-6 and the cycles may accumulate the correlation between the messages during the iterations of detection. This accumulated correlation is difficult to be predicted by mathematical models and can undermine the error performance.
In light of these two aspects, we consider to utilize a neural network (NN) to compensate for the performance loss of the SISO algorithm for FTN detection.

Recently, deep learning supplemented communication systems have shown the potential to further enhance the system's performance \cite{DL_Comm_1,DL_Comm_2,DL_Comm_4,DL_Comm_3, Bryan_ITW2019}. In particular, for the detection and decoding algorithms, the research on autoencoders \cite{Autoencoder_1} and the NN optimization schemes which transform the FGs into NN systems \cite{DLLC, BP_adapt}, has drawn significant interests.
For the autoencoders, an NN system with multiple layers is employed and trained to overcome the issues such as multipath interferences and signal distortions \cite{Autoencoder_1}.
However, since the connection among the multiple layers of the NN model does not rely on the mathematical models of the channel, the NN usually needs a large number of training samples. Generally, more than $2^K$ samples are needed to converge to a good performance \cite{DLSTB}, where $K$ is the information sequence length.
On the other hand, the "unfolded" NN detection or decoding algorithms take advantage of the well-developed channel models \cite{DLLC}, which lead to specific NN connections. However, the flexibility of the NN designs is neglected. Such an NN can only optimize the performance based on a specific graph, which may not lead to the globally optimized performance.
Other than the autoencoders, NN aided algorithms for Turbo equalization systems have also been reported in the literature.
In \cite{TENN}, a deep NN (DNN) was proposed for Turbo equalization by assuming that the log-likelihood ratios (LLRs) of the extrinsic information from the decoder follow a consistent Gaussian distribution with a variance computed by the inverse $J$-function as introduced in \cite{EXIT}.
However, the DNN in \cite{TENN} is constructed by a fully-connected network that has inputs of both the decoder's extrinsic information and the channel information, which indicates that the extrinsic information from the decoder is fed into the NN separately from the channel information. Consequently, this may largely increase the number of neurons in the NN, leading to a high training complexity.

In this paper, we propose an NN assisted approach to compensate the performance loss for the SISO algorithm (referred to as the sum-product detection algorithm (SPDA) thereafter) proposed in \cite{SISO} for detecting coded FTN signals.
We modify the FG of the SPDA by connecting an arbitrary NN to the variable nodes (VNs) via additional function nodes (FNs), where tunable parameters over the edges of the FG are optimized through training.
The proposed algorithm is constructed from a neural network system and it can be trained off-line. Therefore, we name it as deep learning assisted sum-product detection algorithm (DL-SPDA). The inaccuracy of the messages over the edges of the conventional FG due to the limited number of ISI taps and short cycles is expected to be compensated by the NN.
The main contributions of this paper are summarized as follows.
\begin{itemize}
\item We propose a novel DL-SPDA for FTN detection in a modified FG, where an arbitrary NN is concatenated to the original FG of the FTN signaling. We also introduce a new message updating rule to facilitate the application of the proposed DL-SPDA in Turbo equalization receivers. Consequently, the proposed DL-SPDA can be directly applied to a Turbo equalization without considering any particular channel decoding algorithms.
\item We propose a new compatible training technique for NN that is specifically designed for Turbo equalization receivers. The compatible training technique combines the channel information and the extrinsic information from the decoder before training. The extrinsic information is sampled from a consistent Gaussian distribution, where the variance is computed by the inverse $J$-function \cite{EXIT}. 
As a result, the NN is more robust to the variations of extrinsic information from the actual decoder while the training complexity is acceptable.
\item Besides demonstrating that the proposed DL-SPDA can obtain a BER performance gain compared to the conventional FG based FTN detection algorithm, we show the computational complexity analysis of the proposed NN structure.
\item We also derive the theoretical bit error rate (BER) performances for convolutionally encoded FTN systems in order to demonstrate the effectiveness of the proposed DL-SPDA. Specifically, we focus on the case where the detector is imperfect in the sense that only a limited number of ISI taps is considered. By applying approximations, we show that the derived analytical BERs are accurate even for a short block with 250 bits. Meanwhile, the derived analytical BERs also verify that the proposed DL-SPDA approaches the maximum-likelihood (ML) performance.
\end{itemize}
%

%
%

The rest of this paper is organized as follows. In the next section, we will briefly introduce the system model and review the sum-product detection algorithm proposed in \cite{SISO}. Then, in Section \Romannum{3}, we will detail the proposed DL-SPDA, including the new FG model, the modified message updating rules and the compatible training method. The performance analysis for the convolutionally encoded FTN system is discussed in Section \Romannum{4}. The numerical results for both the proposed detection algorithm and the analytical estimates are depicted in Section \Romannum{5} and lastly, the concluding remarks are given in Section \Romannum{6}.
\color{black}



\begin{figure}[t!]
	\centering
	\includegraphics[width=130mm]{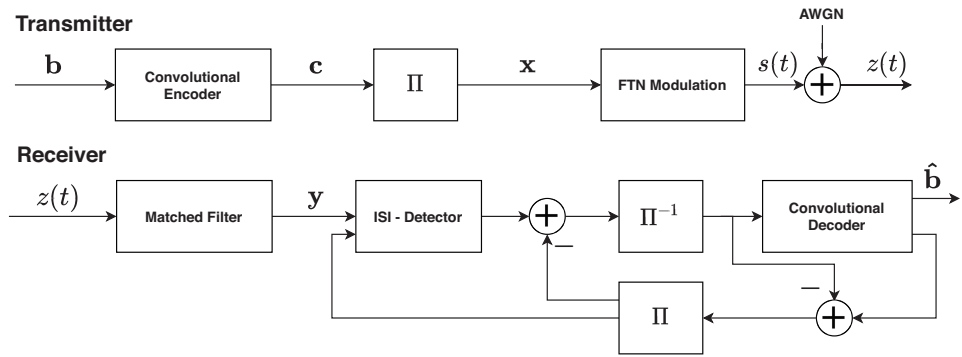}
	\caption{System model.}
    \label{TE}
\end{figure}

\section{Preliminaries}
\subsection{System model}
Without loss of generality, the considered coded FTN system model is shown in Fig. \ref{TE}. Let $\bm{b}$ denote the binary source data with length $K$.
At the transmitter, $\bm{b}$ is convolutionally encoded, resulting in a binary codeword $\bm{c}$ of length $N$.
A sequence of $N$ binary phase-shift keying (BPSK) symbols $\mathbf{x} = [x_1, x_2, ..., x_N]^\text{T}$ is generated after interleaving the bits in $\bm{c}$ with $x_i = (-1)^{c_i}$ for $i \in \{1,..., N\}$.
The transmitted FTN signals are linear modulation signals of the form $s(t) = \sum_n{x_n}{h(t-n\tau T)}$, where $\tau$ is the time acceleration factor of the FTN signaling \cite{FTN_Mazo} and $h(t)$ is a $T$-orthogonal root raised cosine pulse with a roll-off factor $\alpha$.

Assume that the channel is modelled by additive white Gaussian noises (AWGN) with a zero mean and a variance of $\sigma^2$.
The received sequence $\bm{y}$ after a matched filtering and FTN rate sampling is given by $\bm{y} = \mathbf{G}\mathbf{x} + \bm{\eta}$, where $\mathbf{G}$ is a Toeplitz generator matrix, as shown below,
\begin{equation}
\bf{G}=
\begin{pmatrix}
{{g_0}} & {{g_{1}}} & \cdots & {{g_{N-1}}}\\
{{g_{-1}}} & {{g_0}} & \cdots & {{g_{N-2}}}\\
\vdots & \vdots & \ddots & \vdots \\
{{g_{1-N}}} & {{g_{2-N}}} & \cdots & {{g_0}}
\end{pmatrix}.
\end{equation}
The generator matrix consists ISI taps ${g_i} = \int_{ - \infty }^\infty  {h\left( t \right){h^*}\left( {t - i\tau T} \right){\rm{d}}t} $. Specifically, we define $L$ as the number of channel responses with significant energy. The rest ISI taps with insignificant energy are therefore negligible and then set to zeros for simplicity, i.e., ${g_i} = 0$, for $|i| >L$.
Meanwhile, the term $\bm{\eta}$ represents the colored noise samples whose autocorrelation matrix is $\mathbb{E}[\bm{\eta} {\bm{\eta}}^{\rm{H}}] = \sigma^2\mathbf{G}$.

Once the sequence $\bm{y}$ is observed, the receiver performs the Turbo equalization,
where the extrinsic information from the detector and decoder is exchanged iteratively via the interleaver $\mathbf{\Pi}$ or the deinterleaver ${\mathbf{\Pi} ^{ - 1}}$
until the maximum iteration number is reached. The sequence \(\hat {\bm{b}}\) as the estimate of $\bm{b}$ is generated after the Turbo equalization iterations, which is regarded as the output for the receiver.

\subsection{Sum-product detection algorithm}
In this subsection, we briefly review the SPDA proposed in \cite{SISO}. Given the received sequence $\bm{y}$, the SPDA factorizes the $a \ posterior$ probabilities (APPs) $P(\bm{x}|\bm{y})$ of the transmitted sequence $\bm{x}$ mainly based on three types of FNs:


$\bullet$ $O_i(x_i)$ for $i \in \{1,...N\}$: The $a \ priori$ probability for symbol $x_i$ being transmitted.

$\bullet$ $T_i(x_i)$ for $i \in \{1,...N\}$: The symbol likelihood function for symbol $x_i$ being transmitted based on the received symbol $y_i$.

$\bullet$ $I_{i,j}(x_i, x_j)$ for $j \in \{1,...N\}$ and $i \in \{j,...N\}$: The FN that conveys the messages from node $i$ to the interfering node $j$.

The functions of $T_i(x_i)$ and $I_{i,j}(x_i, x_j)$ are defined as \cite{SISO}:
\begin{gather}\label{T}
T_i(x_i) = \text{exp}\bigg{[}\frac{1}{\sigma^2}\text{Re}\bigg{\{}y_ix_i^*-\frac{{\bf{G}}_{i,i}}{2}|x_i|^2\bigg{\}}\bigg{]},\vspace{-5mm}
\vspace{-3mm}
\end{gather}
\begin{gather}\label{I}
I_{i,j}(x_i, x_j) = \text{exp}\bigg{[}-\frac{1}{\sigma^2}\text{Re}\big{\{}{\bf{G}}_{i,j}x_ix_j^*\big{\}}\bigg{]},
\end{gather}
where $x_i^*$ refers to the conjugate of the symbol $x_i$, $\text{Re}\{ \cdot \}$ represents the function that returns the real part of a value, and ${\bf{G}}_{i,j}=g_{i-j}$ is the ${(i-j)}$-th ISI tap. It is derived in \cite{SISO} that $P(\bm{x}|\bm{y}) \propto \prod_{i=1}^{N}\bigg{[}O_i(x_i)T_i(x_i)\prod_{j<i}I_{i,j}(x_i,x_j) \bigg{]}$.

\begin{figure}[t!]
	\centering
	\includegraphics[width=90mm]{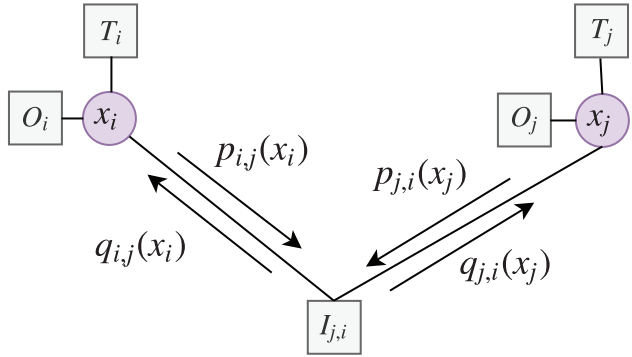}
    \vspace{-7mm}
	\caption{Message updating in the factor graph of SPDA.}
    \label{SPDA_Graph}
\vspace{-4mm}
\end{figure}

Define $q_{i,j}(x_i)$ as the message from the FN $I_{i,j}$ to the VN $x_i$, $p_{i,j}(x_i)$ as the message from the VN $x_i$ to the FN $I_{i,j}(x_i,x_j)$, $o(x_i)$ as the message from the VN $x_i$ to the FN $O_i$, and $Q_i(x_i)$ as the product of all messages incoming to the VN $x_i$, respectively. Here, $Q_i(x_i)$ indicates the proportional probability of the (approximated) APP $P(x_i|\bm{y})$ \cite{SISO}.
An example of part of the FG is given in Fig. \ref{SPDA_Graph}, where the messages are updated according to \cite{SISO}:
\begin{gather}\label{Q_update}
Q_i(x_i) = O_i(x_i)T_i(x_i)\prod_{j \neq i}{q_{i,j}(x_i)},
\end{gather}
\begin{gather}\label{o_update}
o_i(x_i) = \frac{Q_i(x_i)}{O_i(x_i)},
\end{gather}
\begin{gather}\label{p_update}
p_{i,j}(x_i) = \frac{Q_i(x_i)}{q_{i,j}(x_i)},
\end{gather}
\begin{gather}\label{q_update}
q_{i,j}(x_i) = \sum_{x_j}I_{i,j}(x_i, x_j)p_{j,i}(x_j)
\end{gather}

The messages $\{p_{i,j}\}$ and $\{q_{i,j}\}$ are initialized to the same positive values and the messages are updated iteratively until the maximum number of iterations is reached.
Note that the detection complexity of the SPDA is linear to the degree of VNs, i.e., the number of the FNs $I_{i,j}(x_i)$ linked to each VN $x_i$.
Therefore, the SPDA provides a trade-off between the detection performance and detection complexity by intentionally choosing the degree $L_E$ of VNs, where the VN degree $L_E \le  L$ corresponds to the number of ISI taps considered by the SPDA. On the other hand, as we will discuss in Section \Romannum{4}, the coded FTN system's performance is restrained by the number of ISI responses considered by the detector. If only a small number of ISI responses is considered by the detector, an error floor may occur in the high signal-to-noise ratio (SNR) region due to the existence of residual ISI \cite{SY_WJ_FTN}.

It should also be noted that the short cycles contained in the FG will accumulate the correlation between the messages during the iterative update, which may undermine the overall error performance \cite{MCT}. Therefore, to compensate for the potential performance degradation, we propose a deep learning assisted SPDA, the details of which will be discussed in the following section.

%
%
%

\section{Deep Learning Assisted Sum-Product Detection Algorithm}
In this section, the proposed DL-SPDA is introduced.
The basic idea of the proposed DL-SPDA is to transform the SPDA into an NN with additional tuneable multiplicative weights and neuron FNs nested to the VNs.
More specifically, we modify the FG by connecting an NN FN on top of the original FG and then unfold the modified FG into an NN system for training.
Note that the NN FN is directly connected to all VNs.
Therefore, the influence of the residual ISI and short cycles can be compensated by the NN.
Furthermore, we also modify the message-passing rules to make the DL-SPDA be suitable for Turbo equalization receivers. This allows us to train the DL-SPDA without the prior knowledge of the decoder.
Moreover, we propose a compatible training method to improve the performance of the DL-SPDA for Turbo equalization receivers. With the compatible training method, the NN is optimized in terms of the mutual information between the extrinsic information and the transmitted sequence with an acceptable training complexity.

\color{black}

\subsection{New FG model and modified message updating rule}
\begin{figure}[t!]
	\centering
	\includegraphics[width=150mm]{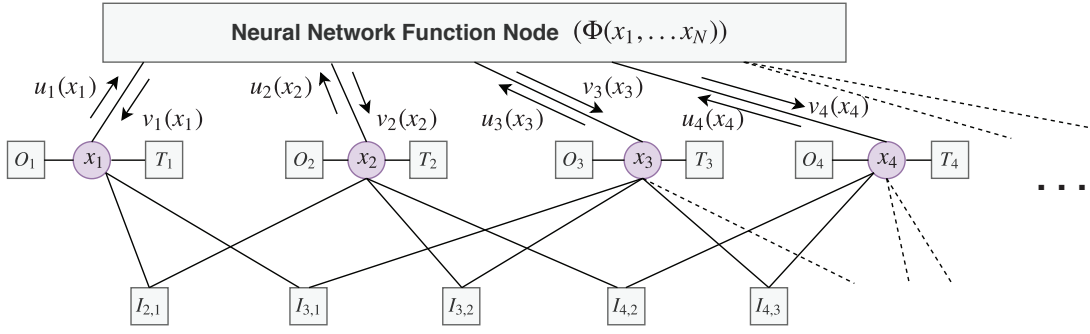}
    \vspace{-6mm}
	\caption{Deep learning assisted sum-product algorithm with $L_E$ = 2.}
    \vspace{-3mm}
    \label{DL-SPDA}
\end{figure}

Conventional NN assisted detection or decoding algorithms introduce tuneable multiplicative weights to the FG and then unfold the message-passing algorithm to an NN system for training and optimization \cite{DLLC}. In terms of the SPDA, trainable weights can be attached to the messages $p_{i,j}(x_i)$ in the corresponding FG, so the updating rule in (\ref{q_update}) is modified as:
\begin{gather}\label{q_update_new}
q_{i,j}(x_i) = \sum_{x_j}I_{i,j}(x_i, x_j)\varsigma_{j,i}p_{j,i}(x_j),
\end{gather}
where $\varsigma_{j,i}$ is the weight attached to the message $p_{j,i}(x_j)$ in the corresponding FG. 
During the off-line training, since $I_{i,j}(x_i, x_j)$ is the interfering node that is irrelevant to the message $p_{j,i}(x_j)$, $I_{i,j}(x_i, x_j)\varsigma_{j,i}$ is treated as the trainable parameters in the neural network. 
Training with the additional weights has shown to improve the sum-product algorithm's performance in a high SNR region \cite{DLLC}.
However, merely tuning the additional weight $\varsigma_{j,i}$ attached to the message $p_{j,i}(x_j)$ will not change the connections in the FG. Therefore, the performance improvement by attaching tuneable weights to the NN is limited.
To this end, we propose to connect an NN FN $\Phi(x_1,...,x_N)$ to the VNs in the FG to compensate for the effects of the residual ISI responses and the correlation induced along the short cycles.
As shown in Fig. \ref{DL-SPDA}, different from the traditional FG, we nest an NN FN to the VNs $x_i$ of the FG, for $i \in \{1,...,N\}$.
The aim of nesting an NN to the FG is two-fold:

$\bullet$ The NN connects to all the VNs. It is expected that all the ISI components among the VNs are considered by the NN.

$\bullet$ The correlation induced during the iteration is expected to be compensated by the NN. The APPs computation for all the VNs in each iteration can be optimized after tuning the parameters in the NN.

Define $u_i(x_i)$ as the message from the variable node $x_i$ to the FN $\Phi(x_1,...,x_N)$ and $v_i(x_i)$ as the message from the FN $\Phi(x_1,...,x_N)$ to the variable node $x_i$ as shown in Fig. \ref{DL-SPDA}.
The conventional sum-product algorithm sums all the intrinsic information for each variable node before passing the extrinsic information to the FN for further processing. This indicates that in a conventional sum-product algorithm, $u_i(x_i)=O_i(x_i)T_i(x_i)\prod_{j \neq i}{q_{i,j}(x_i)}$.
However, in terms of Turbo equalization, the extrinsic information from the decoder will be passed to the detector.
Therefore, the NN needs to be trained with respect to the specific decoder, i.e., the decoding algorithm becomes part of the NN, so that the global optimality can be obtained.
Nevertheless, optimizing the NN which consists of both the detector and decoder has two major problems.
Firstly, the training complexity will be largely increased if the decoder is also included by the NN layers.
Secondly, the optimization of the NN needs to consider the specific channel decoding algorithms, which is inflexible from the design perspective, and undermines the generality of the ISI detector.
Therefore, in order to deal with the extrinsic information from the decoder for Turbo equalization, without introducing any extra off-line training complexity, we propose a new message updating rule for the messages to be passed to the neural network as follows:
\begin{gather}\label{u_update}
u_{i}(x_i) = \prod_{j \neq i}q_{i,j}(x_i).
\end{gather}

Correspondingly, the accumulated APPs for each symbol becomes:
\begin{gather}\label{Q_update_new}
Q_i(x_i) = O_i(x_i)T_i(x_i)v_i(x_i)\prod_{j \neq i}{q_{i,j}(x_i)}.
\end{gather}

The message passed from each VN $x_i$ to the NN FN does not include the messages from $O_i(x_i)$ or $T_i(x_i)$, only the message from the FN $I_{i,j}(x_i,x_j)$ is passed to the NN FN.

According to the proposition, the APP $Q_i(x_i)$ is given by:
\begin{gather}\label{Q_update_new}
Q_i(x_i) = O_i(x_i)T_i(x_i)v_i(x_i)\prod_{j \neq i}{q_{i,j}(x_i)}.
\end{gather}
Compared with the APP update rule in (\ref{Q_update}), (\ref{Q_update_new}) contains the message $v_i(x_i)$ from the NN to the VN $x_i$. Since in each iteration of the Turbo equalization, the $priori$ information $T_i(x_i)$ is constant, the messages from $T_i(x_i)$ will not be passed to the NN for further processing.
We expect the new message updating rule tunes the APPs during each iteration according to the output messages $v_i(x_i)$ of the NN FN.
We summarize the DL-SPDA as follows:

\begin{enumerate}[Step 1:]
  \item Update all the $a \ posterior$ probabilities $\{Q_i\}$ as in (\ref{Q_update_new});
  \item Update the messages $\{p_{i,j}\}$ from the VN $x_i$ to the FN $I_{j,i}$, as in (\ref{p_update});
  \item Update the messages $\{q_{i,j}\}$ from the FN $I_{j,i}$ to the VN $x_i$, as in (\ref{q_update});
  \item Update the messages $\{u_i\}$ from the VN $x_i$ to the NN FN $\Phi(x_1,...,x_N)$, as in (\ref{u_update});
  \item Compute the messages $\{v_i\}$ based on the trained $\Phi(x_1,...,x_N)$;
  \item If the maximum number of iterations is not reached, then go back to Step 1;
\end{enumerate}

\subsection{DL-SPDA with a convolutional NN and its training procedure}
As introduced in Section. \Romannum{2}, the generator matrix of the FTN system is a Toeplitz matrix, which follows a convolutional structure. To explore the convolutional structure of the FTN signaling, we propose to employ a simplified convolutional NN (CNN) to be performed as the NN FN to assist the SPDA \cite{CNN,CNNDetection, CNNRNN_Detection}.
CNNs are widely used in image recognition systems \cite{CNN}.
Traditional CNNs usually involve several convolutional layers (Conv) and max-pooling layers.
The convolutional layer performs the convolution operation of the filters (kernels), where the filters convolve and stride over the input. The max-pooling layer performs downsampling to reduce the spatial size of the convolved features.
A dense layer is appended after the max-pooling layer to provide possibly nonlinear functions \cite{CNN}. 
In \cite{CNNDetection}, a pure CNN based detection algorithm was proposed, where both the max-pooling layers and the dense layer are removed to reduce the training complexity, but a large number of convolutional layers and filters are kept.

\begin{figure}[t!]
	\centering
	\includegraphics[width=110mm]{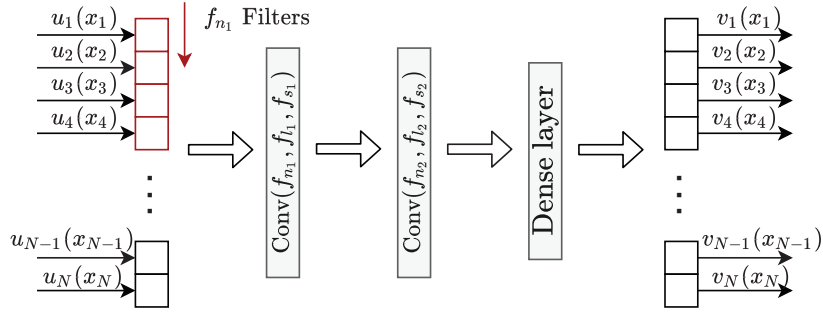}
    \vspace{-5mm}
	\caption{Simplified CNN function node $\Phi(x_1,...,x_N)$.}
    \label{CNN}
\vspace{-6mm}
\end{figure}

In this paper, for the sake of reducing the training complexity of the additional NN FN, we remove the max-pooling layer and simplify the CNN to have only two convolutional layers and one dense layer. The convolutional layer explores the features of the NN FN input message $u_i(x_i)$.
The dense layer is appended to process the features and then delivers the output to the VNs.
%
%
As shown in Fig. \ref{CNN}, the first convolutional layer has $f_{n_1}$ filters and each filter has a size of $ f_{l_1}$, where $f_{l_1}$ indicates that $f_{l_1}$ messages are considered by the filter during each convolutional step.
Let $f_{s_1}$ denote the stride of the sliding window in the first convolutional layer. 
With the attachment of zero padding, the output dimension of the first convolutional layer reduces to $ \lceil N/f_{s_1}\rceil$.
The filter convolves with the messages from $u_1(x_1)$ to $u_N(x_N)$ to compensate any detrimental effect in the message passing that may degrade the detection performance.
A similar operation of filtering is performed in the second convolutional layer which has a corresponding hyper-parameter set of $(f_{n_2}, f_{l_2}, f_{s_2})$.
The output of the convolutional layer is sent to the dense layer to combine and process the filters' results before delivering the messages to the VNs.
The dense layer processes all the filters' results and outputs the message $v_i(x_i)$ for $i \in \{1,...,N\}$.
The negative of a rectified linear activation function (ReLU) is used as the activation function in the dense layer \cite{ReLU}. During each iteration, the CNN has initial weights and biases randomly generated from a truncated normal distribution with a standard deviation of $\sigma_{\text{CNN}_1}$ and  $\sigma_{\text{CNN}_2}$ for the first and second CNN layers, respectively.
For each iteration, messages $u_i(x_i)$ will be updated by using (\ref{u_update}) then passed to the CNN. 
Messages are sent back from the CNN to join the APPs accumulation according to (\ref{Q_update_new}).
By unfolding the FG including the NN FN to an NN system with multiple layers, the overall detector can be trained and optimized.

%
A loss function in an NN training is defined as the errors between the ground truth label and the estimated output.
The cross-entropy loss function has been widely used for the classification problems \cite{DL_Textbook} to minimize the errors between two probability distributions.
By considering the BPSK constellation, we define $\hat{\mathcal{D}}_i^m(x_i)$ as the LLR of the VN $x_i$ at the $m$-th iteration and $m_{max}$ as the maximum number of iterations of DL-SPDA.
The cross-entropy function can be defined as $\mathcal{F}_{ce}(\bm{\mathcal{D}}, \bm{\hat{\mathcal{D}}}^m) = -\frac{1}{N}\sum_{i=1}^{N}\big{(}\mathcal{D}_{x_i}\text{log}(\frac{1}{1+e^{-\hat{\mathcal{D}}^m_{x_i}}}) + (1-\mathcal{D}_{x_i})\text{log}(1-\frac{1}{1+e^{-\hat{\mathcal{D}}^m_{x_i}}})\big{)}$, where $\bm{\mathcal{D}}$ is the ground-truth label of the transmitted bits.
The multi-loss function introduced in \cite{DLLC} and \cite{Syndrome_based} can also be setup to train the tuneable parameters in the FG, which includes the tuneable weights in (\ref{q_update_new}) and the weights and biases introduced in the CNN.
The final loss function for training the NN is then given by:
\begin{gather}\label{loss_func}
\Lambda = \sum_{m=1}^{m_{max}}\gamma^{m_{max}-m}\mathcal{F}_{ce}(\bm{\mathcal{D}}, \bm{\hat{\mathcal{D}}}^m),
\end{gather}
where $\gamma < 1$ is a discount factor to adjust the loss at each iteration. During the training phase, a batch of random transmitted symbols over a range of SNRs is generated as the samples to train the NN.
Once the NN is fully trained, the DL-SPDA can be employed to the Turbo equalization receiver to perform as an ISI detector by exchanging the extrinsic information with the decoder. The extrinsic information from the decoder is passed to the ISI-detector without further tuning the parameters in the NN.

\subsection{Compatible DL-SPDA Training with Mutual Information Compensation}
We have introduced the proposed DL-SPDA in Section \Romannum{3}.A, where the FG is modified by connecting an NN FN and the message updating rule is changed accordingly.
However, in our training phase, we have not considered the extrinsic information passed from the specific decoder. In other words, the extrinsic information to be conveyed to the DL-SPDA at the $\rho$-th iteration of the Turbo equalization is assumed to be an all-zero sequence, i.e. $\mathbf{\Theta}^\rho$ = $\mathbf{0}$.
It can be observed that the DL-SPDA is optimized in the sense of ML detection. Nevertheless, it is well-known that the maximum \emph{a posterior} (MAP) detection outperforms the ML detection for coded systems. Therefore, we propose a compatible training technique which considers the extrinsic information from the decoder.

Inspired by the training technique introduced in \cite{TENN}, we are motivated to deliver the expected distribution of the extrinsic information from the decoder to the NN by assuming that the extrinsic information follows a consistent Gaussian distribution \cite{Anderson_BestCC}, where the mean and variance can be computed from the inverse $J$-function as introduced in \cite{EXIT}.
Note that in \cite{TENN}, the DNN treats the extrinsic information from the decoder as separate input neurons to the NN, i.e., the channel information and the extrinsic information from the decoder are fed into the NN separately. However, this largely increases the number of neurons in the NN and overloads the training process of the whole NN. Therefore, we propose to combine the LLRs of the channel information and the extrinsic information before sending the LLRs to the DL-SPDA, such that the DL-SPDA uses the combined information for further training.

Let $\mathbf{\Psi}^{(\cdot)}$ and $\mathbf{\Upsilon}^{(\cdot)}$ be the LLR realizations for the channel information and the extrinsic information to be passed to the detector, respectively.
Note that the DL-SPDA accepts both $\mathbf{\Psi}^{(\cdot)}$ and $\mathbf{\Upsilon}^{(\cdot)}$ as the inputs.
Each realization of the channel information $\mathbf{\Psi}^{(\cdot)}$ can be generated by transmitting a known data sequence $\mathbf{x}$.
On the other hand, to generate the extrinsic information $\mathbf{\Upsilon}^{(\cdot)}$, we adopt the idea from the previous work in \cite{TENN}.
We assume that the extrinsic information $\mathbf{\Theta}^\rho$ follows a consistent Gaussian probability density function, i.e.,
$\mathbf{\Theta}_i^\rho \sim \mathcal{N}\big{(}(-1)^{x_i} \sigma_E^2/2,\sigma_E^2\big{)}$ for $i \in \{1,...N\}$, where the variance $\sigma_E^2$ can be computed by using the inverse $J$-function \cite{EXIT}, given the mutual information $\mathcal{M}(\mathbf{\Theta}^\rho, \mathbf{x})$ between the extrinsic information $\mathbf{\Theta}^\rho$ and the transmitted sequence $\mathbf{x}$.
To maintain the generality of the NN, we assume that the mutual information $\mathcal{M}(\mathbf{\Theta}^\rho, \mathbf{x}))$ is uniformly ranged between $ [0, 1]$.
In practice, it is sufficient to consider only a set of mutual information $\Omega = \{0, 1/(|\Omega|-1), ..., 1\}$, where $|\Omega|$ is the cardinality of $\Omega$.
Thus, a set of variances can be then determined as $\Xi = \{J^{-1}(\partial), \ \text{for} \ \partial \in \Omega\}$.
Based on the variance set $\Xi$ and the known transmitted sequence $\mathbf{x}$ in the training phase, we are able to randomly generate a realization of the LLR sequence $\mathbf{\Upsilon}^{(\cdot)}_i \sim \mathcal{N}\big{(}(-1)^{x_i} \sigma_E^2/2,\sigma_E^2\big{)}$, for $i \in \{1,...,N\}$.

\begin{figure}[t!]
	\centering
	\includegraphics[width=105mm]{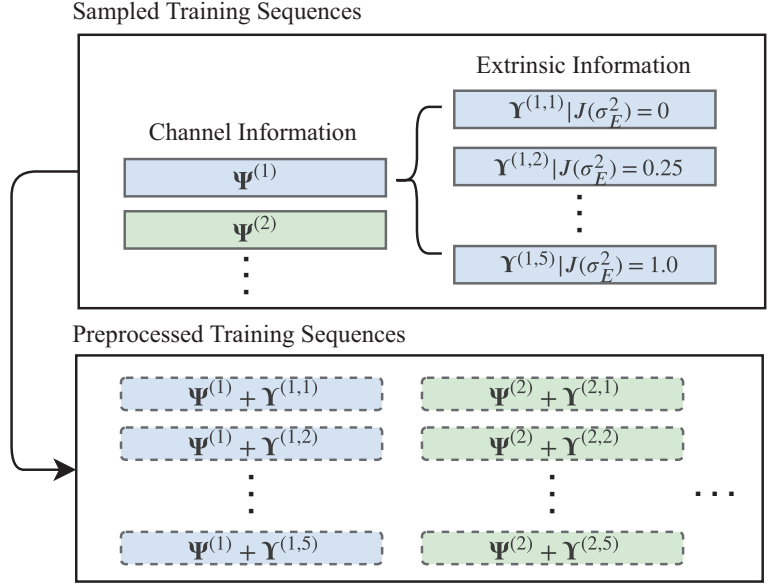}
    \vspace{-4mm}
	\caption{Graphical illustration for the compatible training technique.}
    \label{Compatible_Training}
\vspace{-4mm}
\end{figure}

Fig. \ref{Compatible_Training} illustrates the concept of the compatible training technique to improve the compatibility of the DL-SPDA in the sense of Turbo equalization, where the process is divided into two stages, namely the generation stage and the preprocessing stage.
In Fig. \ref{Compatible_Training}, as an example, $\Omega$ is set to be $\Omega=\{0, 0.25, 0.5, 0.75, 1\}$ and one LLR realization of the extrinsic information is randomly sampled by a specific variance $\sigma_E^2 \in \Xi$ for each realization of the channel information.
Note that during the training phase, based on each variance $\sigma^2_E \in \Xi$, a factor $\mathcal{V}$ can be preset so that $\mathcal{V}$ realizations of the extrinsic information $\mathbf{\Upsilon}^{(\cdot)}$ are randomly sampled based on one realization of the channel information.
This results in that $\mathcal{V} \times |\Omega|$ realizations of the extrinsic information will be randomly sampled based on one realization of the channel information.
We call the stage above as the generation stage.

As for the preprocessing stage, we combine the extrinsic information and the channel information. Specifically, let $\mathbf{\Upsilon}^{(i,j)}$ be the $j$-th realization of the extrinsic information for the
$i$-th realization of the channel information $\mathbf{\Psi}^{(i)}$, for $j \in \{1,..., \mathcal{V} \times |\Omega|\}$.
The message that is passed to the VNs is the combined information sequence, where the elements in $\mathbf{\Psi}^{(i)}$ and $\mathbf{\Upsilon}^{(i,j)}$ are simply added together.
As Fig. \ref{Compatible_Training} demonstrated, each $i$-th realization of the channel information is separately added by the sampled extrinsic information $\mathbf{\Upsilon}^{(i,j)}$ for $j \in \{1,..., 1 \times |\Omega|\}$.
%
%
After all the realizations have been preprocessed, the batch of sequences are fed into the NN for further training.
%

%

%
It can be seen that compared to the training method in Section \Romannum{3}.B, the compatible training technique includes two more stages for processing.
%
This indicates that the input training sequences $\{\mathbf{\Psi}^{(1)} + \mathbf{0},\mathbf{\Psi}^{(2)}+ \mathbf{0},...\}$ in the previous sections are now changed to $\{\mathbf{\Psi}^{(1)}+\mathbf{\Upsilon}^{(1,1)},...,\mathbf{\Psi}^{(1)}+\mathbf{\Upsilon}^{(1,|\Omega|)},\mathbf{\Psi}^{(2)}+\mathbf{\Upsilon}^{(2,1)},...\}$.
These two stages are the preprocessing of the NN training.
It should be noted that the preprocessing does not add any extra input neurons to the NN while it provides the \emph{priori} information for the training.
Therefore, a better error performance can be expected for the DL-SPDA with the compatible training than that without the compatible training.

\section{Performance Analysis for Convolutionally Encoded FTN system}
%
%
As we previously discussed, the DL-SPDA is expected to reduce the detrimental effect of the residual ISI responses and the accumulated correlations of the detecting messages in the FG.
%
%
Therefore, in this section, we aim to theoretically derive the error performance of coded FTN systems in order to evaluate the error performance of the proposed DL-SPDA.
In specific, we focus on the BER performance of the convolutionally encoded FTN systems with a finite block length, where the detection algorithm considers a limited number of ISI taps.
%


%
%
As the generation of the FTN signaling follows the convolutional structure, which can be interpreted by a trellis \cite{Anderson_FTN}.
For the detection of FTN signaling, we define an \emph{error event} as a segment of a path on the FTN trellis that diverges/remerges from/to the correct path \cite{ShuLinBook}.
For an FTN system with a block length of $N$, multiple error events can happen.
Specifically, suppose that the output of a correct path on the FTN trellis for a legitimate transmitted sequence is denoted by $\mathbf{x}$ and the output of the erroneous path is another legitimate transmitted sequence denoted by $\mathbf{x}'$.
Then, for BPSK modulation, we define an error sequence as $\mathbf{e} = \mathbf{x} - \mathbf{x}'$, where $\mathbf{e} \in \mathcal{S}^{(N \times 1)}$ and $\mathcal{S}=\{0,+2, -2\}$.
%
According to the derivations in \cite{Rusek_ML, SY_WJ_FTN}, we find the following definitions and lemma are useful.

\begin{defi}
\emph{(Normalized Euclidean distance)}:
Let $E_b$ be the average energy per information bit. The normalized Euclidean distance $d^2(\mathbf{e})$ of the error sequence $\mathbf{e}$ is defined as
$d^2(\mathbf{e}) \triangleq \frac{1}{2E_b}\mathbf{e}^\text{H}\mathbf{G}\mathbf{e}$.
\end{defi}
\begin{defi}
\emph{(Operative Euclidean distance \cite{SY_WJ_FTN})}:
Denote by $\mathbf{F}$ the truncated version of $\mathbf{G}$, where $\mathbf{F}$ is a Toeplitz matrix constructed by circular shift of the vector $\{0,...,0,g_{-L_E},...,g_0,...,g_{L_E},0,...,0\}$.
The operative Euclidean distance $d_{\text{ope}}^2(\mathbf{e}) $ of the error sequence $\mathbf{e}$ is defined as
$d_{\text{ope}}^2(\mathbf{e}) \triangleq \frac{1}{2E_b}\mathbf{e}^{\rm{H}}\mathbf{F}\mathbf{e}$.
\end{defi}
\begin{defi}
\emph{(Hamming error weight)}:
The Hamming error weight $w$ is defined as the number of non-zero elements of an error sequence $\mathbf{e}$.
\end{defi}
\begin{lemma}
\emph{(Error probability for $\mathbf{e}$ \cite{Ungerboeck})}:
Given the transmitted sequence $\mathbf{x}$, the probability of the ML detector declares the sequence $\mathbf{x}' = \mathbf{x} + \mathbf{e}$ as the detection output is given by
$P_{\mathbf{e}} = Q(\sqrt{\frac{d^2(\mathbf{e})E_b}{N_0}})$.
\end{lemma}
In the following, we will evaluate the ML performance of coded FTN system with a short block length (finite-length). In particular, we consider truncated ISI responses (finite-taps) and approximately full ISI responses (full-taps).
The approximation of full ISI responses means that we consider $L$ most recent ISI taps as described in Section \Romannum{2}.A.

\subsection{Finite-Length Full-Taps Coded FTN system}
To begin with, we start from the ML estimation of the \textbf{uncoded} FTN system.
%
%
%
Let us consider that the error sequence $\mathbf{e}$ has a Hamming error weight $w$.
Therefore, it is obvious that there are in total $\mathcal{A}=\binom{N}{w}2^w$ combinations for $\mathbf{e}$.
For the convenience of notation, we use $\mathbf{e}^{(w,j)} \in \mathcal{S}^{(N \times 1)}$ to denote a specific error sequence $\mathbf{e}$ with a Hamming error weight $w$, where $j \in \{1,...,\mathcal{A} \}$ indicates the $j$-th instance of all possible combinations of $\mathbf{e}$.
Then, based on Lemma 1, the BER $P_b$ can be upper-bounded by applying the union bound \cite{ShuLinBook}, such as:
\begin{align}
P_b < \sum\limits_{w = 1}^N {{}}{\sum\limits_j^{\rm \mathcal{A}} \frac{w}{{2^wN}}{Q\Bigg{(}\sqrt {\frac{{{d^2(\mathbf{e}^{(w,j)})}{E_b}}}{{{N_0}}}} \Bigg{)}} } \label{Pb_FLFull_UB},
\end{align}
where
\begin{align}\label{Q_Func}
Q(\alpha)=\int\limits_\alpha ^\infty  {\frac{1}{{\sqrt {2\pi } }}\exp ( - {\rho ^2}/2)d} \rho,
\end{align}
is the well-known Q-function \cite{ShuLinBook}.
%
%
%
%

We now consider the \textbf{convolutionally encoded} FTN system, where a uniform interleaver is assumed to be applied between the convolutional encoder and the BPSK modulator.
Suppose a rate-$R$ binary terminated convolutional code (CC) $\mathcal{C}'$ is utilized. We define $K_b$ bit streams as the encoder's inputs and $N_b$ bit streams as outputs, i.e., $R=K_b/N_b$.
According to (\ref{Pb_FLFull_UB}), the BER performance can be estimated by considering all possible error sequences, i.e., the distance spectrum.
However, conventional methods, which assume that the all-zero sequence is transmitted, to estimate the BER performance of CC over AWGN channels are not applicable \cite{ShuLinBook}. This is because the Euclidean distance between coded FTN signals can be different even for error sequences with the same Hamming error weight \cite{Li2019code}.
For the ease of presentation, we refer to the error sequence $\mathbf{e} = \mathbf{x} - \mathbf{x}'$ after the FTN modulation as the \textbf{FTN error sequence}, and refer to the binary error vector of the difference between two legitimate CC codewords as the \textbf{CC error sequence}.
It is obvious that given the pair of CC codewords, the FTN error sequence and CC error sequence share the same Hammming error weight.
Therefore, we aim to list all the possible CC error sequences and estimate the BER according to the corresponding FTN error sequences.

In the following, we consider that only a \textbf{single} error event is occurred over the CC trellis. Note that this simplification is widely considered in the literature \cite{ShuLinBook}.
Therefore, we intend to find out all the possible legitimate FTN error sequences that are induced by a single error event in the CC trellis.
Without loss of generality, we consider CC codewords with information bit length $K'$, which is long enough to contain all the possible codewords
covering the CC error events with a relatively small Hamming error weight $w$.
Define $\mathbf{c},\mathbf{c}' \in \mathcal{C}'$ as the instances of the $2^{K'}$ legitimate codeword.
By listing all the possible codewords, we are able to find the CC error sequences corresponding to each single error event $\varepsilon$ and $\mathbf{c}$.
We use $U(\mathbf{c}, \varepsilon) = \{\varepsilon | \mathcal{I}_{\varepsilon} = i, \mathcal{L}_{\varepsilon} = l, \mathcal{W}_{\varepsilon} = w \}$ to denote the set of possible single error events for a legitimate codeword $\mathbf{c}$, where each error event $\varepsilon$ in $U(\mathbf{c}, \varepsilon)$ has an event error length $\mathcal{L}_{\varepsilon} = l$ and the corresponding CC error sequence has Hamming error weight $\mathcal{W}_{\varepsilon} = w$ between $\mathbf{c}$ and $\mathbf{c}'$, corresponding to $\mathcal{I}_{\varepsilon} = i$ different information bits.

As a uniform interleaver is assumed to be employed between the convolutional encoder and the FTN modulator, each possible FTN error sequence corresponding to the CC error sequence is therefore, equally likely to be considered. Define $\mathbf{e}^{(w,o)}|\varepsilon$ as the $o$-th legitimate FTN error sequence with a Hamming error weight $w$ for a given CC event error $\varepsilon$ and $\mathcal{A}_{\varepsilon}$ as the total number of legitimate FTN error sequences for a given CC error event $\varepsilon$.
The BER for the convolutionally encoded FTN system can now be upper bounded by:
\begin{align}\label{Pb_FLFullCoded_FTN}
{P_b} \lesssim \sum\limits_{w = d_{\text{min}}}^N\sum\limits_{\mathbf{c}}^{\mathcal{C}'}\sum\limits_{\varepsilon}^{U(\mathbf{c}, \varepsilon)}\sum\limits_o^{\mathcal{A}_{\varepsilon}} \frac{{{ {\frac{\mathcal{I}_{\varepsilon}}{{NR}}{NR/K_b-\mathcal{L}_\varepsilon+1\choose 1}Q\Bigg{(}\sqrt {\frac{{{d^2\big{(}\mathbf{e}^{(w,o)}|\varepsilon\big{)}}{E_b}}}{{{N_0}}}} \Bigg{)}} } }}{\mathcal{A}_{\varepsilon}2^{K'}},
\end{align}
It can be seen that the denominator part in (\ref{Pb_FLFullCoded_FTN}) is due to the assumptions of binary symmetric inputs of the convolutional encoder and the employment of the uniform interleaver between the convolutional encoder and the FTN signaling.

\subsection{Finite-Length Finite-Taps Coded FTN system}
With the analysis of the finite length coded FTN system which considers full ISI taps, we can now investigate the error performance for the ML estimate when a finite number of ISI taps $L_E < L$ is considered by the detector.
As in the previous section, we firstly consider the case of \textbf{uncoded} finite-taps FTN signaling. Following \cite{SY_WJ_FTN} and Definition 2,
we intend to find the error probability $P_{\mathbf{e}}^{'}$ under the condition that a finite number of ISI responses is considered at the receiver, which is equivalent of finding the probability of $P(\text{Re}\big{\{} \mathbf{e}^\text{H}\bm{\eta} + \mathbf{e}^\text{H}(\mathbf{G}-\mathbf{F})\mathbf{x}\big{\} > E_bd^2_{\text{ope}}(\mathbf{e})})$.
It can be seen that $\mathbf{e}^\text{H}(\mathbf{G}-\mathbf{F})\mathbf{x}$ represents the influence of the residual ISI that is not considered by the matrix $\mathbf{F}$.
Define the term $\wp \triangleq \text{Re}\big{\{} \mathbf{e}^\text{H}\bm{\eta} + \mathbf{e}^\text{H}(\mathbf{G}-\mathbf{F})\mathbf{x}\big{\}}$.
In particular, the term $\wp$ can be modeled as a Gaussian variable according to the central limit theorem and it can be shown that $\mathbb{E}[\wp] = 0$ and the variance $\mathbb{E}[\wp\wp^{\rm{H}}] = N_0 E_b d^2(\mathbf{e})+2E_b\sigma^2_{\text{R},\mathbf{e}}$, where $\sigma^2_{\text{R},\mathbf{e}}$ refers to the variance of the term $\mathbf{e}^\text{H}(\mathbf{G}-\mathbf{F})\mathbf{x}$ induced by the residual ISI taps for a given error sequence $\mathbf{e}$. In specific, we have $\sigma^2_{\text{R},\mathbf{e}} = \frac{1}{2E_b} \mathbb{E}[\mathbf{x}^\text{H}(\mathbf{G}-\mathbf{F})\mathbf{e}\mathbf{e}^\text{H}(\mathbf{G}-\mathbf{F})^\text{H}\mathbf{x}]$ \cite{SY_WJ_FTN}.
Based on the derivations in \cite{SY_WJ_FTN}, the error probability for an ML detection if finite taps of ISI responses are considered by the detector is approximated by the following lemma:
\begin{lemma}
\emph{(Error probability for $\mathbf{e}$ with finite ISI responses) }:
Given the transmitted sequence $\mathbf{x}$ and assume that the ML detector considers finite ISI responses with a length of $L_E$, the probability of the ML detector declares the sequence $\mathbf{x}' = \mathbf{x} + \mathbf{e}$ as the detection output is given by
${P_{\mathbf{e}}^{'}} \approx Q\Bigg{(}\sqrt {\frac{{{E_b}d_{{\rm{ope}}}^2(\mathbf{e})}}{{{N_0}}}\frac{{d_{{\rm{ope}}}^2(\mathbf{e})}}{{{d^2}(\mathbf{e}) + \frac{{2\sigma _{{\text{R},\mathbf{e}}}^2}}{{{N_0}}}}}} \Bigg{)}$.
\end{lemma}

Finding the expectation over all the possible transmitted sequence $\mathbf{x}$ for a given $\mathbf{e}$ can be computationally prohibitive.
%
In the following, we introduce a theorem that obtains the lower bound $\sigma^2_{\text{R}_L,\mathbf{e}}$ for $\sigma^2_{\text{R},\mathbf{e}}$.
%
%

\begin{theo}
For BPSK modulation, given an error sequence $\mathbf{e} \in \mathcal{S}^{(N \times 1)}$,
define $\mathcal{P}$ as the set of non-zero elements' positions in $\mathbf{e}$,
the normalized variance of the term $\mathbf{x}^{\rm{H}}(\mathbf{G}-\mathbf{F})\mathbf{e}$ is \textbf{lower bounded} by $\sigma^2_{\text{R},\mathbf{e}} \geq \sigma^2_{\text{R}_L,\mathbf{e}} = \frac{1}{2E_b}\big{[}\sum_{j \in \mathcal{P}}x_j \cdot [(\mathbf{G}-\mathbf{F})\mathbf{e}]_j\big{]}^2$.
\end{theo}
\begin{prf}
Please refer to Appendix \ref{ProofTheorem_1} for the details.
\end{prf}

Based on Theorem 1, we further obtain the following corollary.
\begin{corollary}
Any two error sequences $\mathbf{e}$ and $\mathbf{e}'$ with the same normalized Euclidean distance, $d^2(\mathbf{e}) = d^2(\mathbf{e}')$, and  operative Euclidean distance, $d^2_{\text{ope}}(\mathbf{e}) = d^2_{\text{ope}}(\mathbf{e}')$, have identical values of lower bounds $\sigma^2_{\text{R}_L,\mathbf{e}}$ and $\sigma^2_{\text{R}_L,\mathbf{e}'}$, i.e., $\sigma^2_{\text{R}_L,\mathbf{e}}=\sigma^2_{\text{R}_L,\mathbf{e}'}$.
\end{corollary}
\begin{prf}
Please refer to Appendix \ref{ProofCorollary} for the details.
\end{prf}

Based on Lemma 2, we are able to derive the analytical BERs for the coded FTN signaling, where the ML detector considers finite ISI responses.
As previously described, $\sigma _{{\text{R},\mathbf{e}}}^2$ is derived by computing the expectation over all the possible transmitted sequences $\mathbf{x}$, while $\sigma _{{\text{R}_L,\mathbf{e}}}^2$ is determined based on the error sequence $\mathbf{e}$.
It is reasonable to employ $\sigma _{{\text{R}_L,\mathbf{e}}}^2$ to estimate the error probability for $\mathbf{e}$.
Moreover, to simplify the computation complexity, the distance spectrum of the FTN signaling can be searched based on the FTN error events \cite{Shuangyang,ISI_Distance_Spec,Modulated_Coding}.
For instance, given a CC error event $\varepsilon$ and a Hamming error weight $w$, let $\Big{\{}\big{(}D_{1_\varepsilon}, d^2(\mathbf{\tilde{e}}^{(w,1)}|\varepsilon),d^2_{\text{ope}}(\mathbf{\tilde{e}}^{(w,1)}|\varepsilon)\big{)},...,\big{(}D_{M_\varepsilon},d^2(\mathbf{\tilde{e}}^{(w,M)}|\varepsilon),,d^2_{\text{ope}}(\mathbf{\tilde{e}}^{(w,M)}|\varepsilon)\big{)}\Big{\}}$ be the distance spectrum for the CC error event $\varepsilon$, where $D_{o_\varepsilon}$ denotes the corresponding number of error sequences which have the same Euclidean and operative distance as $\mathbf{\tilde{e}}^{(w,o)}$ and $M_\varepsilon$ denotes the total number of considered error sequences.
According to Corollary 1, any two error sequences with the same Euclidean distance and operative Euclidean distance have the same values of $\sigma_{\text{R}_L,\mathbf{e}}^2$.
Therefore, the BER for a convolutionally encoded FTN system with a finite number of ISI responses can be approximated by:
\begin{align}\label{FiniteTaps_coded_BER}
{P_b} &\approx \sum\limits_{w = d_{\text{min}}}^N\sum\limits_{\mathbf{c}}^{\mathcal{C}'}\sum\limits_{\varepsilon}^{U(\mathbf{c}, \varepsilon)}\sum\limits_o^{M_{\varepsilon}} \frac{1}{\mathcal{A}_{\varepsilon}2^{K'}} \frac{\mathcal{I}_{\varepsilon}}{{NR}}{NR/K_b-\mathcal{L}_\varepsilon+1\choose 1}D_{o_\varepsilon} \cdot
\notag
\\
& Q\Bigg{(}\sqrt {\frac{{{E_b}d_{{\rm{ope}}}^2(\tilde{\mathbf{e}}^{(w,o)}|\varepsilon)}}{{{N_0}}}\frac{{d_{{\rm{ope}}}^2(\tilde{\mathbf{e}}^{(w,o)}|\varepsilon)}}{{{d^2}(\tilde{\mathbf{e}}^{(w,o)}|\varepsilon) + \frac{{2\sigma _{{\text{R}_L,(\tilde{\mathbf{e}}^{(w,o)}|\varepsilon)}}^2}}{{{N_0}}}}}}\Bigg{)}
\end{align}


Note that the analytical BER in (\ref{FiniteTaps_coded_BER}) is an approximation of the real BER due to the use of the lower bound of $\sigma^2_{\text{R},\mathbf{e}}$. As will be shown in the numerical results, in a high SNR region, using $\sigma _{{\text{R}_L,\mathbf{e}}}^2$ has achieved a close estimate to the error performance of the ML detection and decoding.
More importantly, the lower bound $\sigma _{{\text{R}_L,\mathbf{e}}}^2$ becomes closer to $\sigma _{{\text{R},\mathbf{e}}}^2$, if the influence of the residual ISI tends to be insignificant.
Furthermore, (\ref{FiniteTaps_coded_BER}) demonstrates that the ML performance is restricted by the number of ISI taps considered by the detection algorithm. If full-taps are considered, then $d^2_{\text{ope}(\mathbf{e})}$ is identical to $d^2(\mathbf{e})$ and the residual ISI term will vanish.
In the following numerical results, we will show that when finite taps are considered by the conventional SPDA, the performance is limited by the residual ISI taps and an error floor occurs in a high SNR region when the noise has less impact to the performance than the residual ISI. On the other hand, simulation results show that the performance of the DL-SPDA approaches to that of the ML detection and decoding which is consistent with our analysis.


\section{Numerical Results}
\subsection{Bit-error-rate performance of the proposed DL-SPDA}
In this section, we evaluate the performance of the proposed DL-SPDA scheme for convolutionally encoded FTN systems.
Without loss of generality, we consider the coded FTN systems
with $\tau=0.5$ and $\tau=0.6$, respectively, where the (7, 5) 4-state rate-1/2 non-recursive CC is used.
At the receiver, Turbo equalization is performed, where the extrinsic information is exchanged between the DL-SPDA detector and the BCJR decoder for CC. Compatible training method is applied to the DL-SPDA.
To have a fair comparison, we also perform the SPDA \cite{SISO} and the channel shortening (CS) method as proposed in \cite{CS}. 
Furthermore, truncated-BCJR detection algorithm \cite{MAP_Covalope} for FTN detection is also evaluated to demonstrate the accuracy of the analytical bound, i.e., (\ref{Pb_FLFullCoded_FTN}) and (\ref{FiniteTaps_coded_BER}), when a limited number of ISI taps is considered by the detection algorithm.
Note that, in order to improve the error perforamnce, the BCJR algorithm usually requires a terminated trellis \cite{ShuLinBook}.
Therefore, in our simulation, we intentionally transmit additional symbols to terminate the ISI trellis.
Specifically, four additional code symbols are transmitted for the BCJR decoder for CC, while $2 \times L_E$ additional symbols are transmitted for the truncated-BCJR algorithm.
The analytical BER of a specific code rate $R$ is summed over the CC minimum Hamming distance $w = d_{\text{min}} = 5$ to a Hamming error weight 8 and the approximate full-taps BERs consider $L = 11$ ISI responses.

The corresponding code rate $R$ with the implementation of the truncated-BCJR detection algorithm and BCJR decoding algorithm is provided in the figures for reference, which is also used for the calculation of analytical BERs.
In this paper, the hyper-parameters to train the NN system are shown in Table \ref{table:1}.
\vspace{-0mm}
\begin{table}[t!]
\centering
\begin{center}
 \begin{tabular}{c| c}
\hline
$\sigma_{\text{CNN}_1}$ & 0.03 \\
 \hline
 $\sigma_{\text{CNN}_2}$ & 0.03 \\
 \hline
 $(f_{n_1}, f_{l_1}, f_{s_1})$ & (3, 8, 5)  \\
 \hline
$(f_{n_2}, f_{l_2}, f_{s_2})$ & (1, 3, 1)\\
\hline
 Optimizer & Root Mean Square Propagation\\
  \hline
 SNR Range (dB) &  [6, 8] \\
 \hline
 Learning Rate & 0.001 \\
 \hline
 Batch per SNR & 60 \\
 \hline
 Compatible Training Factor $\mathcal{V}$ & 12 \\
 \hline
 $\Omega$ & \{0.2, 0.4, 0.6, 0.8, 1\} \\
 \hline
 $\gamma$ & 0.9 \\
 \hline
 $m_{max}$ & 6 \\
 \hline
\end{tabular}
\end{center}
\caption{Hyper-parameters for training the DL-SPDA.}
\label{table:1}
\vspace{-10mm}
\end{table}

Note that for a BPSK modulation, the LLR values can be computed by $u_i(x_i)$ for $i \in \{1,..., N\}$, where the LLR values serve as the input of the CNN FN.

\begin{figure}[t!]
	\centering
	\includegraphics[width=90mm]{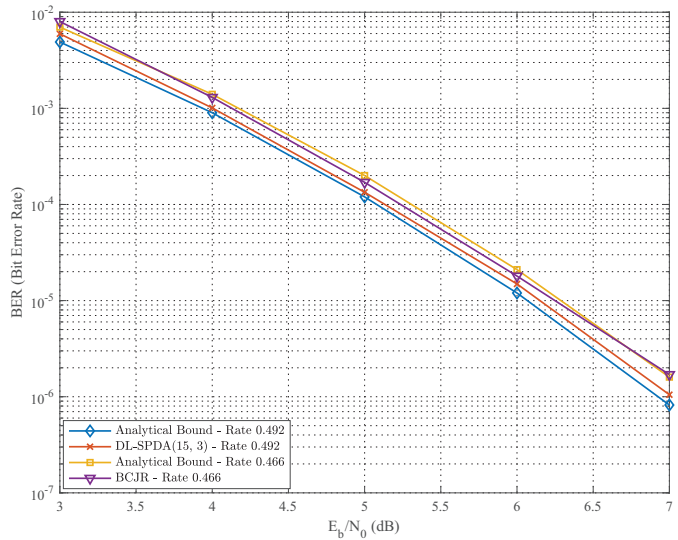}
    \vspace{-6mm}
	\caption{The BER of FTN signaling with $\tau = 0.6$, $\alpha = 0.3$, CC(7, 5) and $N$ = 250.}
    \label{N250_Bound_Performane}
\vspace{-7mm}
\end{figure}

\begin{figure}[t!]
	\centering
	\includegraphics[width=90mm]{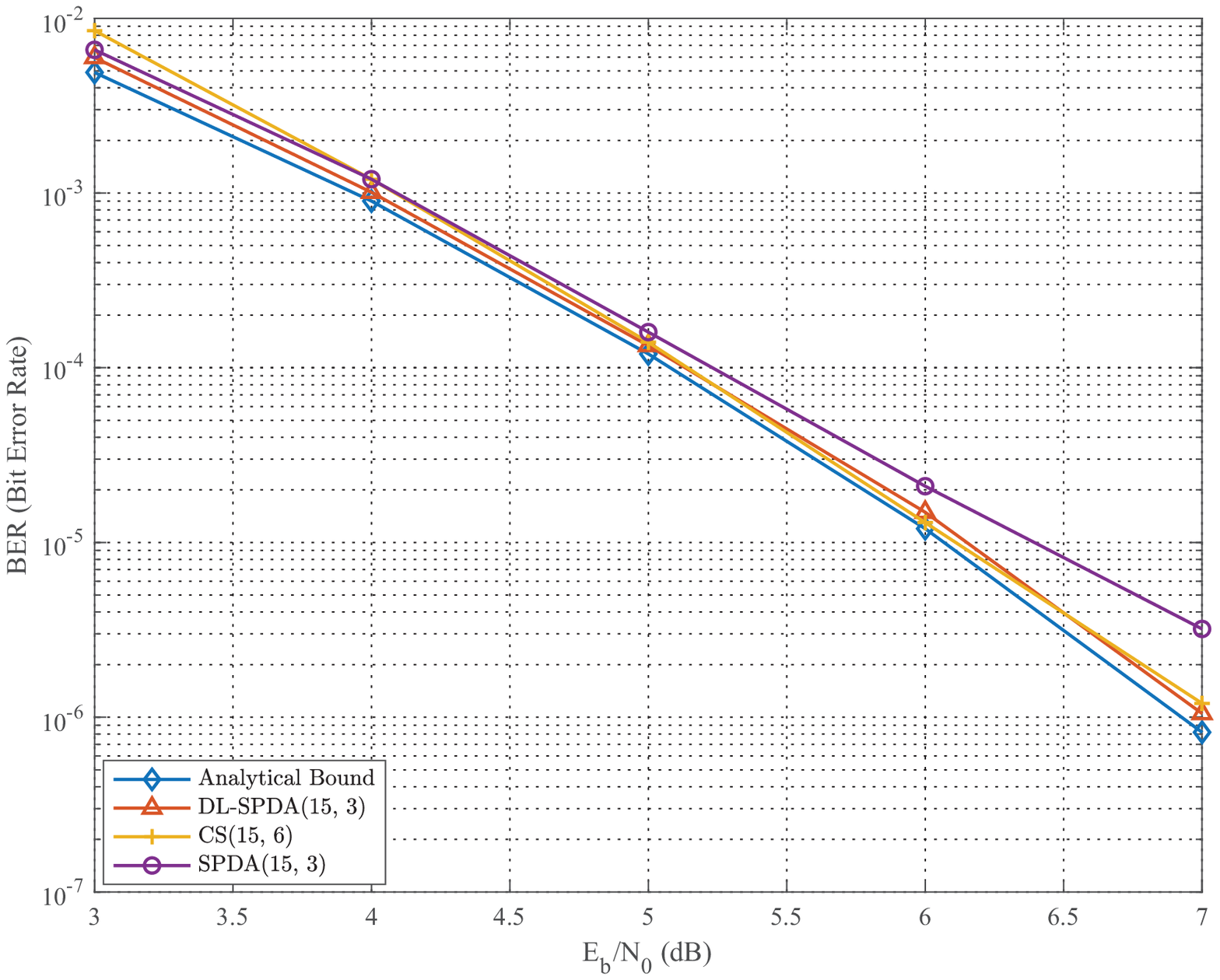}
    \vspace{-6mm}
	\caption{The BER of FTN signaling with $\tau = 0.6$, $\alpha = 0.3$, CC(7, 5) and $N$ = 250.}
    \label{N250_Tau06}
\vspace{-7mm}
\end{figure}

\begin{figure}[t!]
	\centering
	\includegraphics[width=90mm]{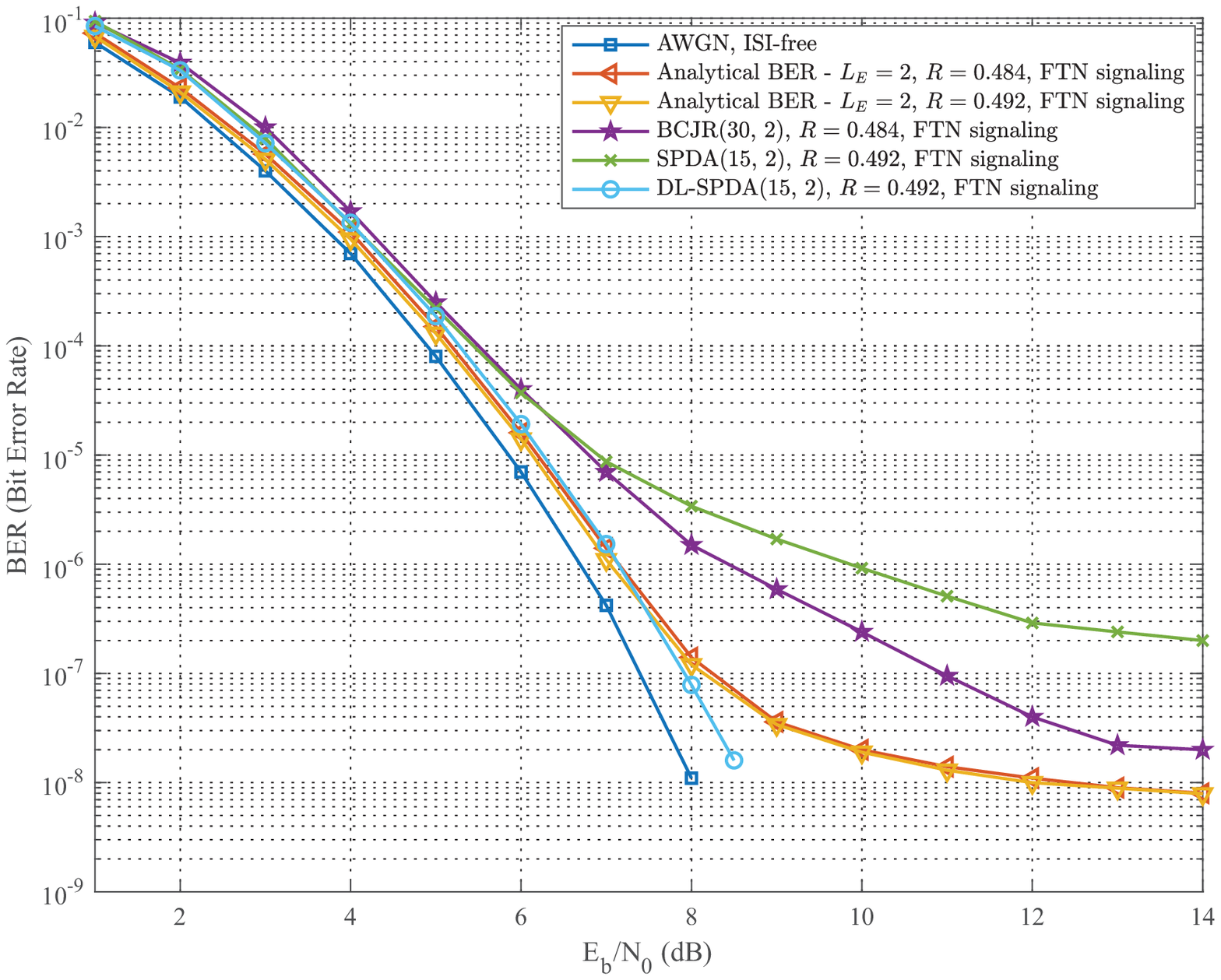}
    \vspace{-6mm}
	\caption{The BER of FTN signaling with $\tau = 0.6$, $\alpha = 0.3$, CC(7, 5) and $N$ = 250.}
    \label{N250_Tau06_Trunc2Taps}
\vspace{-7mm}
\end{figure}

\begin{figure}[t!]
	\centering
	\includegraphics[width=90mm]{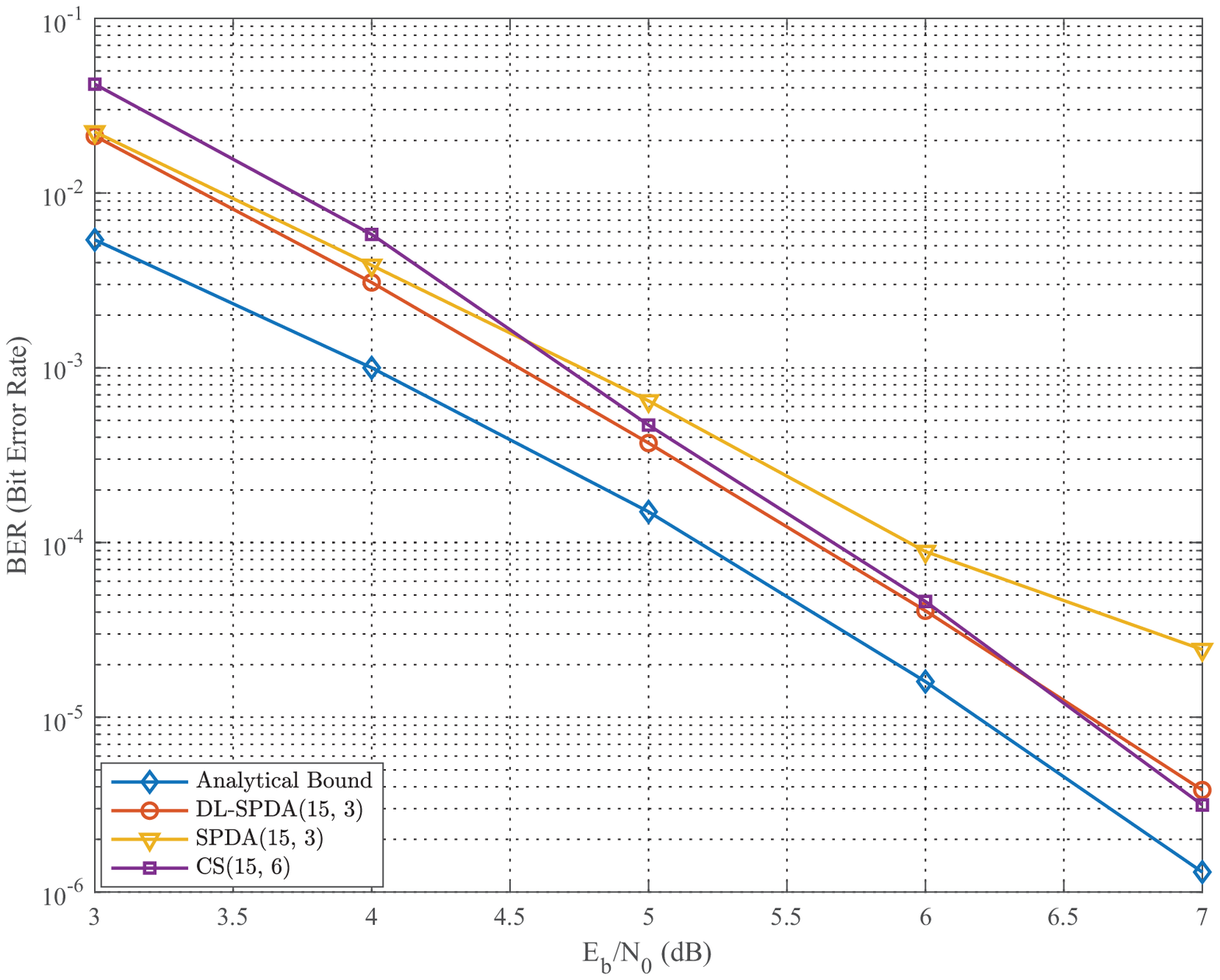}
    \vspace{-6mm}
	\caption{The BER of FTN signaling with $\tau = 0.5$, $\alpha = 0.3$, CC(7, 5) and $N$ = 250.}
    \label{N250_Tau05}
\vspace{-7mm}
\end{figure}

The error performance of various detection algorithms for the FTN systems is shown in Figs.  \ref{N250_Bound_Performane}, \ref{N250_Tau06}, \ref{N250_Tau06_Trunc2Taps} and \ref{N250_Tau05}.
Let $\rho_{max}$ indicate the number of iterations of the Turbo equalization and $L_E$ denotes the number of truncated ISI taps.
DL-SPDA($\rho_{max}$, $L_E$) indicates the proposed DL-SPDA detection method, and BCJR($\rho_{max}$, $L_E$) refers to the corresponding truncated-BCJR detection algorithm \cite{MAP_Covalope}.
Both DL-SPDA and SPDA utilize 6 iterations for updating the messages. Unless specially notified, we compare the error performance at BER $\approx 1 \times 10^{-5}$.

In Fig. \ref{N250_Bound_Performane}, the comparison between analytical bound and the proposed algorithm of an FTN system with $\tau=0.6$ and $N = 250$ is shown, where we can observe that DL-SPDA(15, 3) has achieved a closed performance to the analytical bound with a code rate $R = 0.492$.
Furthermore, we demonstrate the correctness of the derived analytical BERs by performing BCJR(15, 7), where the error performance has a close match to the analytical BERs with a code rate $R=0.466$.

In Fig. \ref{N250_Tau06}, the simulation results of an FTN system with $\tau=0.6$ and $N = 250$ are provided. 
It can be seen that DL-SPDA(15, 3) provides 0.35  dB gain compared to the SPDA(15, 3) at a BER = $3.2 \times 10^{-6}$, where both algorithms consider the same number of ISI taps in the original Tanner graph.
This is due to the fact that after the off-line training, the DL-SPDA algorithm has ``learned'' the residual ISIs which are not considered by the original FG.
Meanwhile, the error performance of CS(15, 6) is also provided in Fig. \ref{N250_Tau06}. As observed from the figure,  both DL-SPDA(15, 3) and CS(15, 6) achieve almost the same error performance that is close to the analytical bound. 
However, the proposed DL-SPDA requires less complexity, which will be discussed in Section \Romannum{5}.B.

In Fig. \ref{N250_Tau06_Trunc2Taps}, we provide the comparison between the analytical BERs for finite-length finite-tap ML estimate and the simulation results for coded FTN signaling. The truncated-BCJR detection algorithm and BCJR decoding algorithm are employed. The codeword length is $N = 250$ and $\tau=0.6$.
In order to demonstrate that an NN FN is capable to capture the interference characteristics throughout the off-line training, we set the DL-SPDA to consider 2 ISI taps based on the Tanner graph and we expect the NN FN learns all the ISI taps leading to an error performance that approachs the analytical bound.
The simulation results in the low SNR region show close performances to the analytical BERs.
Since a lower bound for $\sigma^2_{\text{R},\mathbf{e}}$ is implemented as in (\ref{FiniteTaps_coded_BER}), the BERs are slightly lower than the simulation results of the BCJR algorithm in the high SNR region, when 2 taps are considered by the truncated-BCJR algorithm.
It can be seen that the SPDA(15, 2) has an error floor at a BER = $2 \times 10^{-7}$ due to the insufficient considered ISI taps, while with the help of the concatenated neural network, DL-SPDA(15, 2) shows a promising performance with no noticeable error floor at a BER $< 10^{-7}$.

Fig. \ref{N250_Tau05} depicts the error performance of an FTN system with a more difficult ISI, e.g. $N = 250$ and $\tau=0.5$.
The DL-SPDA(15, 3) outperforms SPDA(15, 3) by 0.75 dB at a BER = $2.5 \times 10^{-5}$.
We observe that DL-SPDA(15, 3) has a close performance to the CS(15, 6).
From Fig. \ref{N250_Tau05}, we observe that for conventional SPDA, the performance is poor due to the strong ISI and the short codeword length. 
On the other hand, DL-SPDA can compensate the strong correlation that exists in the detecting messages and help the detector converges to a better performance.

\begin{figure}[t!]
	\centering
	\includegraphics[width=90mm]{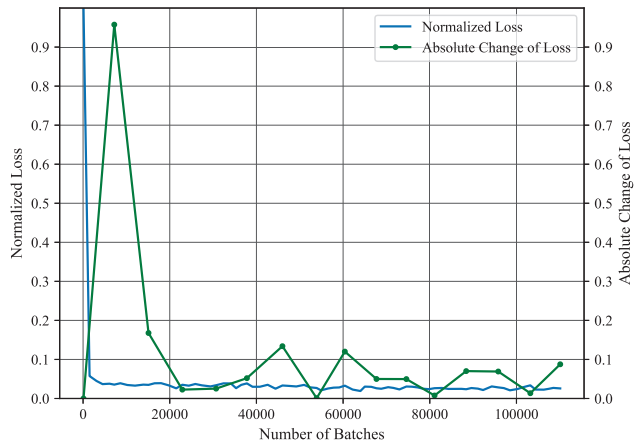}
    \vspace{-6mm}
	\caption{Normalized average loss for every $10^3$ batches of training samples and the absolute change of loss for every $5 \times 10^3$ batches. FTN signaling with $\tau = 0.5$, $\alpha = 0.3$, $N = 250$.}
    \label{Training_Loss}
\vspace{-5mm}
\end{figure}

In Fig. \ref{Training_Loss}, the normalized average training loss for every $10^3$ training batches for an FTN signaling of $\tau = 0.5$ and $N = 250$ with the DL-SPDA(15, 3) is shown.
Define $\xi^a_{\text{avg}}$ as the average loss from $(a-1) \times 5 \times 10^3$ to $a \times 5 \times10^3$ batches ($\xi^0_{\text{avg}} = 1$) and define $\xi^a_{\text{cg}} = |(\xi^a_{\text{avg}}-\xi^{a-1}_{\text{avg}})/\xi^{a-1}_{\text{avg}}|$ as the percentage of the absolute change on the average loss $(\xi^0_{\text{cg}} = 0)$, where $a \in \mathbb{Z}$.
Define that a stable performance of the training is reached after $a \times 5 \times 10^3$ batches, if $\xi^{a'}_{\text{cg}} < 0.1$ for any integer $a' > a$.
From Fig.  \ref{Training_Loss}, we can see that the training phase of the proposed algorithm takes roughly $180 \times 70000$ = $1.26 \times 10^7$ training samples to converge to a relative stable performance.
Benefited from the derived FG and simplified NN model, this number of training samples is much smaller than the conventional NN decoders which generally need more than $2^K$ training samples to converge to a good performance.

\subsection{Computational complexity of the proposed DL-SPDA}
Previous simulation results demonstrate the potential BER performance gain that can be acquired by the proposed DL-SPDA based on the structure of conventional SPDA.
In this section, we discuss the computational complexity of the proposed DL-SPDA.
We mainly compare the computational complexity with the SPDA and the maximum \emph{a posteriror} (MAP) detection algorithm, i.e., log-BCJR algorithm \cite{BCJR_Complexity}.
Compared to the conventional SPDA, DL-SPDA has extra computational complexity introduced by the additional neural network function node. The multiplicative weights attached to the interference node $I_{i,j}\{x_i, x_j\}$ and the interference node itself can be considered as one parameter during the message computation. Therefore, no additional complexity is added by the multiplicative weights of the messages $p_{j,i}(x_j)$ in Section \Romannum{3}.
In Table \ref{Complexity_table}, we show the approximated computational complexity comparison by the number of addition and look-up table accesses, where the operations of comparison, multiplication and non-linear process are assumed to be executed through the look-up table access.
According to \cite{BCJR_Complexity}, the log-MAP algorithm has a computational complexity that grows exponentionally with the considered number of ISI taps.
Comparatively, the SPDA has a linear complexity with the number of ISI per message-passing iteration.
In Table \ref{Complexity_table}, the extra computation complexity that the DL-SPDA requires on top of the SPDA is shown.
For the additions, there are mainly two processes in the DL-SPDA, which are the extra message $v_i(x_i)$ at VN $x_i$ and the process of the neural function node, respectively. Similar to the additions, the processes of the convolutional layer, the ReLu non-linear activation function and the fully connected layer are considered by the look-up table accesses.
In our numerical results, with 6 iterations, DL-SPDA(15, 3) has approximately $2.4 \times 10^5$ additions and $1.0 \times 10^5$ look-up table accesses. 
Comparatively, CS(15, 6) performs linear filtering and the BCJR algorithm. In specific, the BCJR algorithm requires $2.42 \times 10^5$ additions and $1.59 \times 10^5$ look-up table accesses, which is 60$\%$ more than the DL-SPDA algorithm.
Since FTN signaling theoretically introduces infinite number of ISI responses, the difference of computational complexity becomes more significant as more ISI responses are considered by the log-MAP detection algorithm.
More importantly, the proposed algorithm offers the flexility of designing the NN, where the complexity of the proposed algorithm might be further mitigated with advanced NN designs.

\vspace{-0mm}
\begin{table}[h!]
\centering
\begin{center}
 \begin{tabular}{c| c|c}
 \hline
 Operations & Additions & Look-up table accesses \\
\hline
 Log-MAP & $N (15 \times 2^{L_E} + 9)$ & $N(10 \times 2^{L_E} - 4$)\\
 \hline
 SPDA &  $N(32L_E+6)$ & $4N \times L_E$ \\
  \hline
  \multirow{3}{*}{DL-SPDA (extra complexity)}& \multirow{3}{*}{\begin{tabular}{@{}c@{}}$2N + \lceil N/f_{s_1}\rceil (f_{l_1}f_{n_1}+1)+ $ \\ $ \big{\lceil} \lceil N/f_{s_1} \rceil / f_{s_2} \big{\rceil} (f_{l_2}f_{n_1}f_{n_2}+1) + $\\ $\big{\lceil}\lceil N/f_{s_1} \rceil / f_{s_2}\big{\rceil} f_{n_2}N$\end{tabular}}
  & \multirow{3}{*}{\begin{tabular}{@{}c@{}}$\lceil N/f_{s_1}\rceil f_{l_1}f_{n_1} + $\\ $\big{\lceil} \lceil N/f_{s_1} \rceil / f_{s_2} \big{\rceil} (f_{l_2}f_{n_1}f_{n_2}+1) + $ \\ $\big{\lceil}\lceil N/f_{s_1} \rceil / f_{s_2}\big{\rceil} f_{n_2}N$\end{tabular}} \\
  & & \\
  & & \\
 \hline
\end{tabular}
\end{center}
\vspace{-0mm}
\caption{Computational Complexity Comparison Per Iteration.}
\label{Complexity_table}
\vspace{-10mm}
\end{table}

\section{Conclusion}
In this paper, we proposed a DL-SPDA for FTN signaling.
An NN FN is concatenated to the FG of conventional FTN systems to compensate for the detrimental effect of the short cycles in the FG and the residual ISI that are not considered by the FG. The proposed DL-SPDA computes the \emph{a posterior} probability with the assistance of the NN.
A new message updating rule is proposed so that the proposed detection algorithm does not need to be optimized with respect to any particular channel decoder.
Meanwhile, we propose a compatible training technique to improve the compatibility of the DL-SPDA in Turbo equalization.
Moreover, we analysis the finite length coded FTN system's ML BER performance.
The proposed DL-SPDA for coded FTN systems has approached the analytical BERs and the MAP detection and decoding performances.

\appendices

\section{Proof of Theorem 1}
\label{ProofTheorem_1}

We rewrite the residual ISI taps' variance term $\mathbf{x}^\text{H}(\mathbf{G}-\mathbf{F})\mathbf{e}\mathbf{e}^\text{H}(\mathbf{G}-\mathbf{F})^\text{H}\mathbf{x}$ into a format of:
\begin{align}\label{Mtx_e} & \mathbf{x}^\text{H}(\mathbf{G}-\mathbf{F})\mathbf{e}\mathbf{e}^\text{H}(\mathbf{G}-\mathbf{F})^\text{H}\mathbf{x} =[\mathbf{x}^\text{H}(\mathbf{G}-\mathbf{F})\mathbf{e}]^2\notag \\
&= \bigg{[}\sum_{j \in \mathcal{P}}x_j \cdot [(\mathbf{G}-\mathbf{F})\mathbf{e}]_j + \sum_{j \not\in \mathcal{P}}x_j \cdot [(\mathbf{G}-\mathbf{F})\mathbf{e}]_j\bigg{]}^2,
\end{align}
where $\mathcal{P}$ indicates the set of positions of the non-zero elements in $\mathbf{e}$, $[(\mathbf{G}-\mathbf{F})\mathbf{e}]_j$ indicates the $j$-th element of the sequence and $\cdot$ denotes the element-wise multiplication.
The variance of the term $\mathbf{e}^{\rm{H}}\mathbf{G}\mathbf{e}$ induced by the residual ISI taps for a given error sequence $\mathbf{e}$ can now be derived as $\sigma^2_{\text{R},\mathbf{e}} = \frac{1}{2E_b}\mathbb{E}\bigg{\{}\Big{[}\sum_{j \in \mathcal{P}}x_j \cdot [(\mathbf{G}-\mathbf{F})\mathbf{e}]_j + \sum_{j \not\in \mathcal{P}}x_j \cdot [(\mathbf{G}-\mathbf{F})\mathbf{e}]_j\Big{]}^2\bigg{\}}$.

Applying the Jensen's inequality, we have
\begin{align}
\mathbb{E}_{\mathbf{x}|\mathbf{e}}\Big{\{}[\mathbf{x}^\text{H}(\mathbf{G}-\mathbf{F})\mathbf{e}]^2\Big{\}} \geq \big{[}[\mathbb{E}_{\mathbf{x}|\mathbf{e}}(\mathbf{x})]^\text{H}(\mathbf{G}-\mathbf{F})\mathbf{e}\big{]}^2
\end{align}
The second term in the second line of (\ref{Mtx_e}) vanishes as the expectation over the $\{x_j, \ \text{for } \ j \not\in \mathcal{P}\}$ is 0 and the first term is determined on $\mathbf{e}$. Therefore, $\sigma^2_{\text{R},\mathbf{e}} \geq \sigma^2_{\text{R}_L,\mathbf{e}} = \frac{1}{2E_b}\big{[}\sum_{j \in \mathcal{P}}x_j \cdot [(\mathbf{G}-\mathbf{F})\mathbf{e}]_j\big{]}^2$.

\section{Proof of Corollary 1}
\label{ProofCorollary}
Given two error sequence $\mathbf{e}$ and $\mathbf{e}'$ have the same Euclidean distance and operative Euclidean distance, then we have $\frac{1}{2E_b}(\mathbf{e})^\text{H}\mathbf{G}\mathbf{e} = \frac{1}{2E_b}(\mathbf{e}')^\text{H}\mathbf{G}\mathbf{e}'$. Similar to Theorem 1, the equation can be transformed into the element-wise summation by:
\begin{equation}
\frac{1}{2E_b}\sum_{j \in \mathcal{P}_1}e_{j} \cdot [\mathbf{G}\mathbf{e}]_j = \frac{1}{2E_b}\sum_{j \in \mathcal{P}_2}e'_{j} \cdot [\mathbf{G}\mathbf{e}']_j,
\end{equation}
where $\mathcal{P}_1$ and $\mathcal{P}_2$ are the sets of positions of the non-zero elements of $\mathbf{e}$ and $\mathbf{e}'$, respectively. Then, given $\mathbf{F}$ is a truncated matrix of $\mathbf{G}$ that the elements out of the truncation are filled with 0s, we have $\frac{1}{2E_b}\sum_{j \in \mathcal{P}_1}e_{j} \cdot [\mathbf{F}\mathbf{e}]_j = \frac{1}{2E_b}\sum_{j \in \mathcal{P}_2}e'_{j} \cdot [\mathbf{F}\mathbf{e}']_j$, which results in $\frac{1}{2E_b}\sum_{j \in \mathcal{P}_1}\big{(}e^{(1)}_{j} \cdot [\mathbf{G}\mathbf{e}]_j - e^{(1)}_{j}[\mathbf{F}\mathbf{e}]_j\big{)} = \frac{1}{2E_b}\sum_{j \in \mathcal{P}_2}\big{(}e'_{j} \cdot [\mathbf{G}\mathbf{e}']_j - e'_{j} \cdot [\mathbf{F}\mathbf{e}']_j\big{)}$. It can be modified to
\begin{equation}
\frac{1}{2E_b}\sum_{j \in \mathcal{P}_1}\big{(}e_{j} \cdot [\mathbf{(G-F)}\mathbf{e}]_j\big{)} = \frac{1}{2E_b}\sum_{j \in \mathcal{P}_2}\big{(}e'_{j} \cdot [\mathbf{(G-F)}\mathbf{e}']_j\big{)}.
\end{equation}
The error sequence $\mathbf{e}$ and the transmitted sequence $\mathbf{x}$ follow a one-to-one mapping at the non-zero elements' positions $\mathcal{P}_1$ and $\mathcal{P}_2$. Let $\mathbf{x}$ and $\mathbf{x}'$ be the corresponding transmitted sequences for $\mathbf{e}$ and $\mathbf{e}'$, respectively. Then, the corollary is proved that:
\begin{equation}
\frac{1}{2E_b}\big{[}\sum_{j \in \mathcal{P}_1}x_{j} \cdot [(\mathbf{G}-\mathbf{F})\mathbf{e}]_j\big{]}^2  = \frac{1}{2E_b}\big{[}\sum_{j \in \mathcal{P}_2}x'_{j} \cdot [(\mathbf{G}-\mathbf{F})\mathbf{e}']_{j}\big{]}^2
\end{equation}

\label{SecondAppendix}



\end{document}